%% file: aone.tex
\documentclass[12pt]{article}
\usepackage{amsmath}
\usepackage{amssymb}
\usepackage{vector}
\def\eps{\varepsilon}
\font\tencmmib=cmmib10 \skewchar\tencmmib '60
\newfam\cmmibfam
\textfont\cmmibfam=\tencmmib

\def\bbox{\quad\hbox{\vrule \vbox{\hrule \vskip2pt \hbox{\hskip2pt
\vbox{\hsize=1pt}\hskip2pt} \vskip2pt\hrule}\vrule}}
\def\lessim{\ \lower4pt\hbox{$
\buildrel{\displaystyle <}\over\sim$}\ }
\def\gessim{\ \lower4pt\hbox{$\buildrel{\displaystyle >}
\over\sim$}\ }
\def\P{{\cal P}}

\def\eps{{\varepsilon}}
\def\ch{{\mbox{ch}}}

\def\th{{\mbox{th}}}

\def\RS{{\mbox{RS}}}

\def\la{{\Bigl\langle}}
\def\ra{{\Bigr\rangle}}

\def\qed{\hfill\break\rightline{$\bbox$}}
\newcommand{\e}{\mathbb{E}}
\newcommand{\p}{\mathbb{P}}
\newcommand{\Reals}{\mathbb{R}}

\newcommand{\vsi}{{\vec{\sigma}}}
\newcommand{\vrho}{{\vec{\rho}}}

\newcommand{\te}{\mathbb{T}}

\newtheorem{lemma}{Lemma}
\newtheorem{theorem}{Theorem}

\makeatletter
\@addtoreset{equation}{section}

\makeatother

\input invlat.tex

\begin{document}

\title{
Exponential control of overlap in the replica method for  
$p$-spin Sherrington-Kirkpatrick model.}

\author{ 
Dmitry Panchenko\thanks{Department of Mathematics, Massachusetts Institute
of Technology, 77 Massachusetts Ave, Cambridge, MA 02139
email: panchenk@math.mit.edu. This work is partially supported by NSF grant.
}\\
{\it Department of Mathematics}\\
{\it Massachusetts Institute of Technology}\\
}
\date{}

\maketitle

\begin{abstract}
In \cite{TLD} the large deviations limit $\lim_{N\to\infty}(Na)^{-1}\log \e Z_N^a$ 
for the moments of the partition function $Z_N$ in the Sherrington-Kirkpatrick model
\cite{SK} was computed for all real $a\geq 0.$ For $a\geq 1$ this result extends the classical
physicist's replica method that corresponds to integer $a.$
We give a new proof for $a\geq 1$ in the case of the pure $p$-spin SK model
that provides a strong exponential control of the overlap.
\end{abstract}
\vspace{0.5cm}



Abbreviated title: Overlap control in replica method.

\section{Introduction and main results.}

For integer $N\geq 1$ we consider $\Sigma_N=\{-1,+1\}^N$ 
and a Gaussian Hamiltonian (process) $H_N(\vsi)$ indexed
by $\vsi\in \Sigma_N.$ We assume that its covariance satisfies
\begin{equation}
\frac{1}{N}\e H_N(\vsi^1) H_N(\vsi^2) = \xi(R_{1,2}),
\label{correlation}
\end{equation}
where
\begin{equation}
R_{1,2}=\frac{1}{N}\sum_{i\leq N}\sigma_i^1\sigma_i^2
\label{overlap}
\end{equation}
is the overlap of configurations $\vsi^1, \vsi^2$
and $\xi$ is a smooth function such that
\begin{equation}
\xi(0)=0,\, \xi(x)=\xi(-x),\,\xi''(x)>0 \mbox{ if } x>0.
\label{xi}
\end{equation}
In fact, our main results will be obtained in the case when $\xi(x)=\beta^2 |x|^p$
for $p\geq 2.$ When $p\geq 2$ is an even integer, the corresponding Hamiltonian
$$
H_N(\vsi)=\frac{\beta}{N^{(p-1)/2}}\sum_{1\leq i_1,\ldots,i_p \leq N}
g_{i_1,\ldots,i_p}\sigma_{i_1}\ldots \sigma_{i_p}
$$
is called a pure $p$-spin SK Hamiltonian. Here $(g_{i_1,\ldots,i_p})$
are i.i.d. standard Gaussian random variables.
Let us define a function $\theta(x) = x\xi'(x) - \xi(x)$ that due to convexity
assumption on $\xi$ satisfies
\begin{equation}
\Delta(a,b) :=\xi(a)-a\xi'(b)+\theta(b)\geq 0
\,\,\,\mbox{ for all } a,b\in\Reals.
\label{positivity}
\end{equation}
This property will be crucial in the same way it was crucial in all recent
progress in the SK model, because it yields the positivity of error terms 
in Guerra type interpolations and, as a result, allows us to control them.
Given the external field parameter $h\in \Reals,$
we define the partition function by
\begin{equation}
Z_N=\sum_{\vsi\in\Sigma_N}\exp\Bigl(H_N(\vsi)+h\sum_{i\leq N} \sigma_i\Bigr).
\label{partition}
\end{equation}
The main results of our paper are motivated by a problem considered in \cite{TLD}.
The goal of that paper was to identify the following large deviations
limit for the partition function
$$
\P(a) = \lim_{N\to\infty} \frac{1}{N a} \log \e Z_N^a
$$
for all $a\geq 0.$ It was shown that the cases $0\leq a< 1$ and $a\geq 1$
are very different. The first case is of the same nature as the Parisi formula,
proved rigorously in \cite{TP}, that corresponds to $a=0.$ The second case
$a\geq 1$ is the generalization of the so called replica method that
corresponds to integer $a\geq 1.$ In our paper we will only be interested in
this second case and, in particular, the following problem. The main result
for $a\geq 1$ in \cite{TLD}, Theorem 9.4, says that
if $a\geq 1$ then 
\begin{equation}
\P(a) = \max_{q \in [0,1]} \RS(q)
\label{PRSa}
\end{equation}
where
\begin{equation}
\RS(q) =
\log 2 + \frac{1}{2}\Bigl(
\xi(1)-\xi'(q) + (1-a)\theta(q)
\Bigr)
+\frac{1}{a}
\log \e \ch^a(z+h).
\label{RS}
\end{equation}
and where a r.v. $z$ has normal distribution $N(0,\xi'(q)).$
(It is interesting to note that $\RS$ is convex in $a$ by main result in
\cite{Par} which, otherwise, is not at all obvious.)
The proof of (\ref{PRSa}) was based on a beautiful convexity argument
in the spirit of the proof of Ghirlanda-Guerra identities \cite{GG}.
The proof also suggested that, assuming that the supremum of $\RS(q)$ is achieved
at a unique point $q_0,$ the distribution of the overlap under a certain
change of density should be concentrated near $q_0$ if the external field
$h\not = 0$ or, by symmetry, near $\pm q_0$ if $h=0.$ This type of
behavior was witnessed in a weak sense - on average over some small
perturbations of the Hamiltonian $H_N(\vsi).$  In the present paper 
we achieve a strong exponential control of the overlap for pure $p$-spin 
case, $\xi(x)=\beta^2 |x|^p$ for $p\geq 2.$

In order to formulate and motivate the results let us sketch the
starting point of the proof of (\ref{PRSa}) in \cite{TLD} which gives the lower bound 
$\P(a)\geq \sup_{q}\RS(q).$ This is based on Guerra's interpolation. 
Let us consider an interpolating Hamiltonian for $0\leq t\leq 1$ given by
\begin{equation}
H_t(\vsi) = \sqrt{t} H_N(\vsi) + \sqrt{1-t} \sum_{i\leq N} z_i  \sigma_i 
+h\sum_{i\leq N} \sigma_i,
\label{Hint}
\end{equation}
where $(z_i)_{i\leq N}$ are i.i.d. $N(0,\xi'(q)).$ 
Define the partition function as above
$$
Z_t = \sum_{\vsi\in \Sigma_N} \exp H_t(\vsi)
\,\,\,\mbox{ and let }\,\,\,
\varphi(t) = \frac{1}{Na}\log \e Z_t^a.
$$
A standard Gaussian integration by parts then shows that
$$
\varphi'(t) = \frac{1}{2}\Bigl(
\xi(1)-\xi'(q) + (1-a)\theta(q)
\Bigl)
+\frac{1}{2}(a-1) \e'\la \Delta(R_{1,2},q)\ra
$$
where $\la\cdot\ra$ is the Gibbs' average with Hamiltonian $H_t(\vsi)$
(we keep its dependence on $t$ implicit)
and $\e'$ denotes the expectation with the following change of density
$$
\forall f,\,\,\, \e' f = \e \te f
\,\,\,\mbox{ for }\,\,\,
\te = \frac{Z_t^a}{\e Z_t^a}.
$$
Since
$$
\varphi(0) = \log 2+ \frac{1}{a}\log \e \ch^a(z+h)
$$
we get
\begin{equation}
\varphi(1)= \RS(q) + \frac{1}{2}(a-1) \int_{0}^1 \e'\la \Delta(R_{1,2},q)\ra dt.
\label{integint}
\end{equation}
By (\ref{positivity}), $\Delta(x)\geq 0$ 
and we get $\P(a)\geq \RS(q).$ In order to show that this lower
bound is sharp for $q=q_0$ we need to show that the last integral in
(\ref{integint}) is small. The strong control of $\Delta(R_{1,2},q_0)$
along the interpolation $0\leq t<1$ was obtained in \cite{TLD} for $1\leq a\leq 2$
but, as we mentioned above, the matching upper bound for general $a\geq 1$
was given only in a weak sense. Of course, the fact that we have a matching upper bound
implies that the above integral is small and, thus,
\begin{equation}
\lim_{N\to\infty}\int_0^1 \e'\la \Delta(R_{1,2},q_0)\ra dt = 0
\label{temp1}
\end{equation}
By itself this does not exclude large values of
$\e'\la \Delta(R_{1,2},q_0)\ra$
at some exceptional points $t$ but our first simple observation does.

\begin{theorem}
For all $a\geq 2$ and $0\leq t<1$ we have
$$
\lim_{N\to\infty} \e'\la \Delta(R_{1,2},q_0)\ra = 0.
$$
\end{theorem}
{\it Proof.}
A simple computation by Gaussian integration by parts gives
(we omit $q_0$ in $\Delta$ for simplicity of notations)
\begin{eqnarray}
2N^{-1}\frac{\partial}{\partial t}\e'\la \Delta(R_{1,2})\ra
&=&
\e'\la \Delta(R_{1,2})^2\ra + 2(a-2)\e'\la \Delta(R_{1,2})\Delta(R_{1,3})\ra
\label{temp2}
\\
&+&
\frac{1}{2}(a-2)(a-3)\e'\la \Delta(R_{1,2}) \Delta(R_{3,4})\ra
-\frac{1}{2}a(a-1) (\e'\la \Delta(R_{1,2})\ra)^2.
\nonumber
\end{eqnarray}
H\"older's inequality applied either to $\la\cdot \ra$ or $\e'$ implies
$$
\e' \la \Delta(R_{1,2})^2\ra 
\geq 
\e'\la \Delta(R_{1,2})\Delta(R_{1,3})\ra
\geq 
\e'\la \Delta(R_{1,2}) \Delta(R_{3,4})\ra
\geq
(\e'\la \Delta(R_{1,2})\ra)^2.
$$
Since the sum of all coefficients on the right hand side of (\ref{temp2})
is zero and the first two are nonnegative for $a\geq 2,$ the above derivative
is nonnegative so that $\e'\la \Delta(R_{1,2})\ra$ is nondecreasing.
Together with (\ref{temp1}) this proves the result.
\qed

This, however, does not give us control of the overlap at $t=1$
and even for $t<1$ it is still a rather weak statement.  For example,
this does not show that the third moment $|R_{1,2}-q_0|^3$
is of a smaller order than the second moment which is needed in order to carry
out second moment computations and prove central limit theorem for the overlap. 
Below we will formulate a stronger statement.

Let us start with a remark about maxima of $\RS(q).$ It is easy
to check that the critical point condition is 
\begin{equation}
\frac{\e \ch^a(z+h)\th^2(z+h)}{\e\ch^a(z+h)} = q 
\,\,\,\mbox{ where }\,\,\,
z\sim N(0,\xi'(q)). 
\label{critq}
\end{equation}
We do not know how to prove that this equation
has a unique solution, even though numerical observations show
that this seems to be the case. For $a=0,$ Lemma 2.4.8 in \cite{SG}
(Guerra-Latala) proves that such solution is unique by showing that
a function $q\to \e \th^2(z+h)/q$ is decreasing.
This, however, is not always true for $a\geq 1.$  

Below we will prove that the overlap essentially can not take 
values $u\in[-1,1]$ such that $\RS(|u|)< \max \RS(q).$
Under an additional (easy-to-check) assumption that $\RS(q)$ has 
a unique maximum at $q_0$ this will imply that the overlap is strongly
concentrated near $q_0.$ By now standard techniques 
(for example, Sections 2.6, 2.7 in \cite{SG}) one can then carry out 
second moment computations and prove central limit theorem for the overlap.

Let us note that the overlap can only take values
$R_{1,2}=k/N$ for integer $-N\leq k\leq N.$ For simplicity of
notations throughout the paper when we write $R_{1,2}=u \in [-1,1]$
we mean that $R_{1,2}=u_N$ for some sequence $(u_N)$ such that 
$\lim_{N\to\infty} u_N = u.$ We would like to show that (for $t=1$)
\begin{equation}
\e'\la I(R_{1,2} = u)\ra \leq L \exp(-N/L) 
\label{expcontrol}
\end{equation}
if one of the following holds:
\begin{enumerate}
\item{
$h\not = 0$ and either $u<0$ or $u\geq 0$ and
$\RS(u) < \P(a).$
}
\item{$h= 0$ and
$\RS(|u|) < \P(a).$
}
\end{enumerate}
Here and everywhere below $L$ denotes a constant that does not depend on $N.$
We need to separate these two cases because without
external field, $h=0,$ the distribution of overlap is symmetric.
In order to prove (\ref{expcontrol}), we will prove a stronger
statement. For $n\geq 1$ let
$$
{\cal C}_n = \Bigl\{ 
U=(u_{l,l'})_{1\leq l,l'\leq n} \,\,:\,\,
U^T=U\geq 0,\,\,\, u_{l,l'}\in [-1,1] \,\,\mbox{ and }\,\, u_{l,l}=1
\Bigr\}.
$$
Given $U \in {\cal C}_n,$ we will write $\{R_{l,l'}=u_{l,l'}\}$
to denote a set of all spin configurations $\vsi^1,\ldots,\vsi^n$
such that $R_{l,l'}=u_{l,l'}$ for all $1\leq l,l'\leq n.$
In fact, since it will always be absolutely clear from the context,
we will abuse the notations and simply write $U$ to denote 
$\{R_{l,l'}=u_{l,l'}\}.$
We define a product partition function with corresponding
constraints on overlaps by
\begin{equation}
Z_n(U) = \sum_{U}\exp\Bigl(
\sum_{l\leq n} H_N(\vsi^l) + h\sum_{l\leq n}\sum_{i\leq N}
\sigma_i^l
\Bigr).
\label{partconstr}
\end{equation}
Also, for $n=2$ and $u\in[-1,1]$ we will write
$Z_2(u)$ instead of $Z_2(U)$ for a matrix $U\in {\cal C}_2$
such that $u_{1,2}=u.$ Let $n$ be such that
\begin{equation}
n\leq a < n+1.
\label{anone}
\end{equation}
Then we can write
$$
\e'\la I(R_{1,2} = u)\ra 
= 
\frac{1}{\e Z_N^a}\e Z_N^{a-2} Z_2(u)
\leq
\frac{1}{\e Z_N^a}\e \bigl(Z_N^{n-1} Z_2(u)\bigr)^{a/(n+1)},
$$
since the last inequality (for the integrands) is equivalent to
$Z_2(u)\leq Z_N^2.$ Therefore, (\ref{expcontrol}) will follow from
$$
\frac{1}{N}\log \e \bigl(Z_N^{n-1} Z_2(u)\bigr)^{a/(n+1)}
\leq 
\frac{1}{N}\log \e Z_N^a - \frac{1}{L}.
$$
Since each overlap takes at most $2N+1$ values it should be obvious that
the left hand side is equivalent for $N\to\infty$ to
$$
\sup_{W}\frac{1}{N}\log \e Z_{n+1}(W)^{a/(n+1)}
$$
where the supremum is taken over all $W\in {\cal C}_{n+1}$ such that
$w_{1,2}=u.$
Thus, for pure $p$-spin model exponential overlap control in (\ref{expcontrol}) follows
from our main result. 

\begin{theorem}\label{Thmain}
Suppose that $\xi(x)=\beta^2 |x|^p$ for $p\geq 2$ and $u\in [-1,1]$
satisfies one of the conditions in (\ref{expcontrol}). If 
$W\in {\cal C}_{n+1}$ is such that $w_{1,2}=u$ then
\begin{equation}
\lim_{N\to\infty} \frac{1}{Na}\log \e Z_{n+1}(W)^{a/(n+1)}
\leq 
\P(a) - \frac{1}{L}
\label{mainin}
\end{equation}
for some $L>0.$
\end{theorem}
The only part of the proof that uses the specific choice of $\xi(x)=\beta^2 |x|^p$
is Lemma \ref{Pmatrix} below. Generalizing Lemma \ref{Pmatrix} would immediately
yield Theorem \ref{Thmain} for other choices of $\xi,$ for example, mixed
$p$-spin Hamiltonians.

Let us mention that the control of the overlap for pure $p$-spin model provided 
by Theorem \ref{Thmain} generalizes the so called replica method which corresponds
to integer $a=n.$ For completeness let us formulate this well known result.

\begin{theorem}
Suppose that $a$ is integer, $\xi$ satisfies (\ref{xi}) and  $u\in [-1,1]$
satisfies one of the conditions in (\ref{expcontrol}). Then (\ref{expcontrol})
holds.
\end{theorem}

The proof of this result follows by well known techniques, see Section 2.15 in \cite{SG}
or \cite{VHP}. The result is proved in the Appendix A of \cite{VHP} only for $h=0$ 
(the authors attribute the proof to Elliott Lieb) but essentially the same argument 
can be extended to $h\not = 0.$ For additional comments see remark following
Lemma \ref{L2} below.

\section{Proof of Theorem \ref{Thmain}.}

First of all, let us note that we write a limit on the left hand side
of (\ref{mainin}) instead of $\limsup$ because the limit exists.
It follows from a standard superadditivity argument by using Guerra-Toninelli
interpolation as in \cite{GT}. We only mention that a condition $a/(n+1)<1$
is important because it implies that a derivative in Guerra-Toninelli interpolation
has a correct sign.

Suppose that all conditions of theorem are satisfied but 
\begin{equation}
\lim_{N\to\infty} \frac{1}{Na}\log \e Z_{n+1}(W)^{a/(n+1)}=\P(a).
\label{contra}
\end{equation}
It will be convenient to assume that one of the elements in the last
column of $W$ instead of $w_{1,2}$ is equal to $u.$ Suppose that
\begin{equation}
W=\left(
\begin{array}{cc}
U & \vec{u}\\
\vec{u}^T & 1
\end{array}
\right),
\label{WU}
\end{equation}
where $U\in {\cal C}_n,$ $\vec{u}=(u_1,\ldots,u_n)^T$ and one of the coordinates
of $\vec{u}$ is equal to $u.$ The inequality $Z_n(U)^{a/n}\leq Z_N^a$
together with (\ref{contra}) implies that
\begin{equation}
\limsup_{N\to\infty}\frac{n}{Na}\log \e Z_n(U)^{a/n}
\leq
n\P(a)
=
\lim_{N\to\infty} \frac{n}{Na}\log \e Z_{n+1}(W)^{a/(n+1)}.
\label{LUstart}
\end{equation} 
We will first obtain a lower bound for the left hand side of (\ref{LUstart}).
Given a vector $\vec{z}\in\Reals^n$ and 
$\vec{\lambda}=(\lambda_{l,l'})_{1\leq l<l'\leq n}$
we define a function
\begin{equation}
\Phi_n(\vec{z},\vec{\lambda}) = 
\sum_{\eps_1,\ldots,\eps_n = \pm 1}
\exp\Bigl(
\sum_{l\leq n}\eps_l (z_l+h) + \sum_{l<l'} \lambda_{l,l'} \eps_l \eps_{l'}
\Bigr).
\label{Phin}
\end{equation}
Given an $n\times n$ covariance matrix $Q$ let $\xi'(Q)=(\xi'(q_{l,l'}))$
and define
\begin{eqnarray}
\psi(Q,\vec{\lambda}) 
&=& 
\frac{1}{2} \sum_{1\leq l,l'\leq n}
\Bigl(
\xi(u_{l,l'}) - u_{l,l'}\xi'(q_{l,l'})
+\Bigl(1-\frac{a}{n}\Bigr) \theta(q_{l,l'})
\Bigr)
\nonumber
\\
&&
-\sum_{l<l'}\lambda_{l,l'} u_{l,l'}
+
\frac{n}{a}\log \e \Phi_n^{a/n}(\vec{z},\vec{\lambda})
\label{psi}
\end{eqnarray}
where $\vec{z}$ has Gaussian distribution $N(0,\xi'(Q)).$

\begin{lemma}\label{Llower}
If $n\leq a$ then 
\begin{equation}
\liminf_{N\to\infty} \frac{n}{Na}
\log \e Z^{a/n}_n(U) 
\geq 
\sup_{Q} \inf_{\vec{\lambda}} \psi(Q,\lambda).
\label{lower}
\end{equation}
\end{lemma}
{\it Proof.}
Let us define $Z_t(U)$ by replacing $H_N(\vsi^l)$
in the definition of the partition function $Z_n(U)$ 
with Hamiltonians
$$
\sqrt{t} H_{N}(\vsi^l) +\sqrt{1-t}\sum_{i\leq N}z_{i,l} \sigma_i^l
$$
where $(z_{i,l})_{l\leq n}$ are independent copies of $\vec{z}$ for $i\leq N$
and let 
$$
\varphi(t) = \frac{n}{Na}
\log \e Z^{a/n}_t(U). 
$$
By Gaussian integration by parts 
\begin{eqnarray}
\varphi'(t) 
&=& 
\frac{1}{2}\sum_{1\leq l,l'\leq n}
\Bigl(
\xi(u_{l,l'})-u_{l,l'}\xi'(q_{l,l'}) + 
\Bigl(1-\frac{a}{n}\Bigr)\theta(q_{l,l'})
\Bigr)
\nonumber
\\
&+&
\frac{1}{2}\Bigl(\frac{a}{n}-1\Bigr)
\sum_{1\leq l,l'\leq n}\e'\bigl\la
\Delta(R^{l,l'},q_{l,l'})
\bigr\ra.
\label{deran}
\end{eqnarray}
The Gibbs average $\la\cdot\ra$ in the last term is taken over
two copies $\Sigma_N^n \times \Sigma_N^n$ and $R^{l,l'}$ denotes the overlap
between configuration $\vsi^l$ from the first copy and configuration
$\vrho^{l'}$ from the second copy. Since $n\leq a$ and $\Delta\geq 0,$
the last term in (\ref{deran}) is nonnegative and, therefore,
$$
\varphi(1)\geq \varphi(0) +
\frac{1}{2}\sum_{1\leq l,l'\leq n}
\Bigl(
\xi(u_{l,l'})-u_{l,l'}\xi'(q_{l,l'}) + 
\Bigl(1-\frac{a}{n}\Bigr)\theta(q_{l,l'})
\Bigr).
$$
It should be obvious from definitions that for all $\vec{\lambda}$
$$
\varphi(0) \leq
-\sum_{l<l'}\lambda_{l,l'} u_{l,l'}
+
\frac{n}{a}\log \e \Phi_n^{a/n}(\vec{z},\vec{\lambda}).
$$
However, using standard large deviations techniques one can show
that this bound is sharp
$$
\lim_{N\to\infty} \varphi(0) 
=
\inf_{\vec{\lambda}}
\Bigl(
-\sum_{l<l'}\lambda_{l,l'} u_{l,l'}
+
\frac{n}{a}\log \e \Phi_n^{a/n}(\vec{z},\vec{\lambda})\Bigr).
$$
Since the choice of $Q$ was arbitrary, this finishes the proof of Lemma.
\qed

If $a\leq n+1$ then exactly the same proof will produce the upper bound 
for the right hand side of (\ref{LUstart}) because the last term
in (\ref{deran}), with $n$ now replaced by $n+1$, will be negative.
Let $P$ be an $(n+1)\times(n+1)$ covariance matrix and let
$\vec{y}$ be a Gaussian random vector with distribution 
$N(0,\xi'(P)).$ Given $\vec{\gamma}=(\gamma_{l,l'})_{1\leq l,l,\leq n+1}$ 
we define
\begin{eqnarray}
\Psi(P,\vec{\gamma}) 
&=& 
\frac{1}{2}\frac{n}{n+1} \sum_{1\leq l,l'\leq n+1}
\Bigl(
\xi(w_{l,l'}) - w_{l,l'}\xi'(p_{l,l'})
+\Bigl(1-\frac{a}{n+1}\Bigr) \theta(p_{l,l'})
\Bigr)
\nonumber
\\
&&
-\frac{n}{n+1}\sum_{1\leq l<l'\leq n+1}\gamma_{l,l'} w_{l,l'}
+
\frac{n}{a}\log \e \Phi_{n+1}^{a/(n+1)}(\vec{y},\vec{\gamma}).
\label{Psi}
\end{eqnarray}

\begin{lemma}\label{L2}
If $a\leq n+1$ then
\begin{equation}
\frac{n}{Na}\log \e Z_{n+1}(W)^{a/(n+1)}
\leq
\inf_{P}\inf_{\vec{\gamma}}
\Psi(P,\vec{\gamma}). 
\label{upper}
\end{equation}
\end{lemma}
{\it Remark.}
This upper bound was given in Section 8 of \cite{TLD} and
the question answered here (only for $p$-spin model)
was posed as an open problem there. In order to prove Theorem \ref{Thmain},
the first urge is to try to find parameters $P$ and $\vec{\gamma}$ that would
witness (\ref{mainin}). However, this direct approach seems intractable.
In fact, to understand the difficulty, one should look at the simplest case 
of integer $a$ for which the answer is provided by Lieb's argument in \cite{VHP}.
Suppose that $a=n+1.$  Then taking $P=W$ and $\vec{\gamma}=0$ in (\ref{upper}) yields
\begin{eqnarray*}
\frac{1}{N}\log \e Z_{a}(W) 
&\leq&
\log \sum_{\vsi}\exp\Bigl(
\frac{1}{2}\sum_{l,l'}\xi'(w_{l,l'})\sigma_l \sigma_{l'} 
+ h\sum_{l\leq a}\sigma_l
\Bigr)
-\frac{1}{2}\sum_{l,l'}\theta(w_{l,l'})
\\
&=& 
\log \sum_{\vsi}\exp\Bigl(
\sum_{l<l'}\xi'(w_{l,l'})\sigma_l \sigma_{l'} 
+ h\sum_{l\leq a}\sigma_l
\Bigr)
-\frac{1}{2}\sum_{l,l'}\theta(w_{l,l'})
+\frac{1}{2}\sum_{l\leq a}\xi'(1).
\end{eqnarray*}
An ingenious argument in \cite{VHP} then shows that the
supremum of the right hand side over $W$ is achieved on
the diagonal when all $w_{l,l'}=w$ and is strictly
less for $W$ off the diagonal. It is easy to see that
for a constant matrix $W$ the above bound becomes 
$a\RS(w)$ and, as a result, we obtain exponential
control for values of the overlap that do not maximize
$\RS(q).$ We do not see how to extend Lieb's argument for 
non-integer values of $a,$ in particular, because the  
last term in (\ref{Psi}) is much less explicit in $P.$
Our approach will be quite different and the main idea will
be to relate upper and lower bounds of Lemmas 1 and 2.
It will follow from the argument below that, similarly to
the integer case, the bound (\ref{upper}) is always
maximized on the diagonal.
\qed

Lemmas 1 and 2 and (\ref{LUstart}) imply that
\begin{equation}
\sup_{Q}\inf_{\vec{\lambda}}
\psi(Q,\vec{\lambda})
\leq 
\inf_{P}\inf_{\vec{\gamma}}
\Psi(P,\vec{\gamma}).
\label{lup}
\end{equation}
To prove Theorem \ref{Thmain} we need to extract useful
information from comparing these upper and lower bounds.
Let us start by rewriting the first line in (\ref{Psi}).
Recalling (\ref{WU}) and regrouping the terms we can write it as 
$\mbox{I} + \mbox{II} + \mbox{III}$ where
\begin{eqnarray*}
\mbox{I}
&=&
\frac{1}{2}\frac{n}{n+1} \sum_{1\leq l,l'\leq n}
\Bigl(
\xi(u_{l,l'}) - u_{l,l'}\xi'(p_{l,l'})
+\Bigl(1-\frac{a}{n}\Bigr) \theta(p_{l,l'})
\Bigr),
\\
\mbox{II}
&=&
\frac{1}{2}\frac{n}{n+1}
\Bigl(
\xi(1) - \xi'(p_{n+1,n+1})+(1-a)\theta(p_{n+1,n+1})
\Bigr),
\\
\mbox{III}
&=&
\frac{a}{2}\Bigl(\frac{n}{n+1}\Bigr)^2 
\Bigl(\
\theta(p_{n+1,n+1}) +
\frac{1}{n^2} \sum_{1\leq l,l'\leq n}\theta(p_{l,l'})
-\frac{2}{n} \sum_{1\leq l\leq n}
\theta(p_{l,n+1})
\\
&&
\hspace{2.3cm}
+\
\frac{2(n+1)}{na}
\sum_{1\leq l\leq n}
\bigl(
\xi(u_{l}) - u_{l}\xi'(p_{l,n+1}) +\theta(p_{l,n+1})
\bigr)
\Bigr).
\end{eqnarray*}
Terms I and II were defined to match similar terms in the definition of
$\psi$ and $\RS$ and III ensures that the sum of all three gives the first line
in (\ref{Psi}). We would like to choose the matrix $P$ such that
$\mbox{III}\leq 0.$ Unfortunately, at this point we were able to do this
only in the case when $\xi(x)=\beta^2 |x|^p$ and this is the only part
of the proof that uses the specific choice of $\xi$ in Theorem \ref{Thmain}.

\begin{lemma}\label{Pmatrix}
Suppose that $\xi(x)=\beta^2 |x|^p.$ If we take
\begin{equation}
P=\left(
\begin{array}{cc}
s^2 \vec{u}\vec{u}^T & \vec{u}\\
\vec{u}^T & s^{-2}
\end{array}
\right)
\,\,\,\mbox{ where }\,\,\,
s=|\vec{u}|_p^{-1/2}=\Bigl(\frac{1}{n}\sum_{1\leq l\leq n}|u_l|^p\Bigr)^{-1/(2p)}
\label{Pchoice}
\end{equation}
then $P^T=P\geq 0$ and $\mbox{\rm III} = 0.$
\end{lemma}
{\it Proof.} We can write $P=a\,a^T$ for 
$a=(s\vec{u}^T,s^{-1})^T$ and, thus, $(Px,x)=(a^Tx,a^T x)\geq 0.$
Since $\theta(x)=\beta^2(p-1)|x|^p,$
plugging this choice of $P$ into III gives
\begin{eqnarray*}
\frac{2}{a\beta^2(p-1)}\Bigl(\frac{n+1}{n}\Bigr)^2\mbox{III}
&=&
s^{-2p}+
\frac{1}{n^2}\sum_{1\leq l,l'\leq n} s^{2p}|u_l|^p |u_{l'}|^p
-
\frac{2}{n}\sum_{1\leq l\leq n} |u_l|^p
\\
&=&
s^{-2p}
+ s^{2p}|\vec{u}|_p^{2p} - 2|\vec{u}|_p^{p} = 0,
\end{eqnarray*}
with the above (optimal) choice of $s.$
\qed

From now on assume that $P$ is defined by (\ref{Pchoice}).
Let
\begin{equation}
Q = s^2 \vec{u}\vec{u}^T  
=|\vec{u}|_p^{-1} \vec{u}\vec{u}^T 
\,\,\,\mbox{ and }\,\,\,
q=p_{n+1,n+1}=s^{-2}=|\vec{u}|_p.
\label{choices}
\end{equation}
Then the first line in (\ref{Psi}) is equal to $\mbox{I}+\mbox{II}$ where
\begin{eqnarray}
\mbox{I}
&=&
\frac{n}{n+1}\ \frac{1}{2} \sum_{1\leq l,l'\leq n}
\Bigl(
\xi(u_{l,l'}) - u_{l,l'}\xi'(q_{l,l'})
+\Bigl(1-\frac{a}{n}\Bigr) \theta(q_{l,l'})
\Bigr),
\label{one}
\\
\mbox{II}
&=&
\frac{n}{n+1}\ \frac{1}{2}
\Bigl(
\xi(1) - \xi'(q)+(1-a)\theta(q)
\Bigr).
\label{two}
\end{eqnarray}
Let us now look at the second line in (\ref{Psi}).
Let us take $\vec{\gamma}$ such that $\gamma_{l,n+1}=0$ for all $l\leq n$
and let us rename $\gamma_{l,l'}=\lambda_{l,l'}$ for $1\leq l<l'\leq n.$ Then,
obviously,
\begin{equation}
\sum_{1\leq l<l'\leq n+1}\gamma_{l,l'} w_{l,l'}
=
\sum_{1\leq l<l'\leq n}\lambda_{l,l'} u_{l,l'}.
\label{three}
\end{equation}
Given a random vector $\vec{y}$ with distribution $N(0,\xi'(P))$
let $\vec{y}=(\vec{z},z)$ so that $\vec{z}$ has distribution
$N(0,\xi'(Q))$ and $z$ has distribution $N(0,\xi'(q)).$
With the above choice of $\vec{\gamma}$ it should be obvious that
$$
\Phi_{n+1}(\vec{y},\vec{\gamma})
=
\Phi_{n}(\vec{z},\vec{\lambda})\times
2\ch(z+h)
$$
and, therefore, by H\"older's inequality
\begin{equation}
\frac{n}{a}\log \e \Phi_{n+1}^{a/(n+1)}(\vec{y},\vec{\gamma})
\leq 
\frac{n}{n+1}\ \Bigl(\,
\frac{n}{a}\log \e \Phi_{n}^{a/n}(\vec{z},\vec{\lambda})
+
\frac{1}{a}\log \e (2\ch(z+h))^a\Bigr).
\label{Holder}
\end{equation}
Combining this with (\ref{one}), (\ref{two}) and (\ref{three}) proves that
\begin{equation}
\Psi(P,\vec{\gamma}) \leq
\frac{n}{n+1}\ \psi(Q,\vec{\lambda}) 
+\frac{n}{n+1}\ \RS(q)
\label{Psipsi}
\end{equation}
and (\ref{lup}) implies
$$
\inf_{\vec{\lambda}}
\psi(Q,\vec{\lambda})
\leq 
\sup_{Q'}\inf_{\vec{\lambda}}
\psi(Q',\vec{\lambda})
\leq
\inf_{P'}\inf_{\vec{\gamma}} \Psi(P',\vec{\gamma})
\leq
\frac{n}{n+1}\ 
\inf_{\vec{\lambda}}
\psi(Q,\vec{\lambda}) 
+\frac{n}{n+1}\ \RS(q).
$$
Solving this inequality gives
$$
\inf_{\vec{\lambda}}\psi(Q,\vec{\lambda}) 
\leq n\, \RS(q) = n\, \RS(|\vec{u}|_p).
$$
Plugging this back into (\ref{Psipsi}) gives
\begin{equation}
\inf_{\vec{\gamma}}\Psi(P,\vec{\gamma}) 
\leq n\, \RS(|\vec{u}|_p).
\label{big}
\end{equation}
Let us consider two alternatives - either all elements of $\vec{u}$ are equal in
absolute value or not
In the first case $|\vec{u}|_p = |u|$
and Lemma \ref{L2} implies that
$$
\frac{1}{Na}\log \e Z_{n+1}(W)^{a/(n+1)}
\leq \RS(|\vec{u}|_p) = \RS(|u|),
$$
which finishes the proof of Theorem \ref{Thmain} for $h=0$ or $h\not = 0$
and $u\geq 0.$ 

Therefore, it remains to consider the cases when either not
all elements of $\vec{u}$ are equal in absolute value 
or $h\not = 0$ and $u<0.$ The following holds.

\begin{lemma}\label{L4}
If either not all elements of $\vec{u}$ are equal in absolute value 
or $h\not = 0$ and $u<0$  then the inequality (\ref{Holder}) 
is strict.
\end{lemma}

{\it Remark.} For simplicity of the proof, we will use the particular 
choices of $\xi(x)=\beta^2 |x|^p$ and $P$ in (\ref{Pchoice}). However, it
should be easy to generalize the proof for general $\xi.$ Lemma \ref{Pmatrix}
is the only place where the specific form of $\xi$ was essential.

{\it Proof.}
H\"older's inequality in (\ref{Holder}) will be equality only if
\begin{equation}
\Phi_{n}(\vec{z},\vec{\lambda})
=\mbox{const}\cdot\ch(z+h)^n
\label{Hol}
\end{equation}
almost surely. However, since $(\vec{z},z)$ have normal distribution
with covariance $\xi'(P),$ 
\begin{equation}
z_l = a_l z\,\,\,\mbox{ for }\,\,\, a_l = \xi'(s^2)\xi'(u_l).
\label{za}
\end{equation}
For $\vec{\eps}\in \{-1,+1\}^n,$ let  
\begin{eqnarray*}
f_1(\vec{\eps}) 
&=& 
\frac{1}{W_1}
\exp\Bigl(
h \sum_{1\leq l\leq n}\eps_j + \sum_{1\leq l<l'\leq n} \lambda_{l,l'} \eps_l \eps_{l'}
\Bigr),
\\
f_2(\vec{\eps}) 
&=& 
\frac{1}{W_2}
\exp\Bigl(
h \sum_{1\leq l\leq n}\eps_j 
\Bigr)
\end{eqnarray*}
be two probability functions on $\{-1,+1\}^n,$ where $W_1$ and $W_2$
are corresponding normalizing factors.
Recalling the definition of $\Phi_n$ and using (\ref{za}), 
(\ref{Hol}) can be rewritten as 
\begin{equation}
\sum_{\vec{\eps}}
f_1(\vec{\eps})\exp\Bigl(
z\sum_{1\leq l\leq n}a_l \eps_l 
\Bigr) 
=
\sum_{\vec{\eps}}
f_2(\vec{\eps}) \exp\Bigl(
z\sum_{1\leq l\leq n}\eps_l
\Bigr).
\label{f1f2}
\end{equation}
almost surely for $z$ and, since both sides are continuous,
for all $z\in\Reals.$ Letting $z\to \infty$ implies that
$\sum |a_l| = n$ and by (\ref{za}),
$$
\frac{1}{n}\sum_{1\leq l\leq n}|\xi'(u_l)| = \xi'(s^{-2})
=\xi'(|\vec{u}|_p).
$$
Since $\xi'(x)=p \beta^2 |x|^{p-1}\mbox{sgn}(x),$ this is equivalent to
$|\vec{u}|_{p-1} = |\vec{u}|_p$ which can happen only if
all elements of $\vec{u}$ are equal in absolute value.
If not, this proves that the inequality (\ref{Holder}) is strict.
If they are equal, then all $a_l=\pm 1$ and it remains to consider
the case $h\not = 0$ and $u<0.$ Equation (\ref{f1f2}) means that
moment generating functions of $\sum a_l\eps_l$ under the law with
p.f. $f_1$ and of $\sum \eps_l$ under the law with p.f. $f_2$ are
equal and, therefore, their distributions are equal. For example,
$$
\p_1\Bigl(\sum_{1\leq l\leq n}a_l \eps_l=n\Bigr) 
=\p_2\Bigl(\sum_{1\leq l\leq n}\eps_l=n\Bigr).
$$
If we denote $\vec{s}=(\mbox{sgn}(a_1),\ldots,\mbox{sgn}(a_n))$
and $\vec{1}=(1,\ldots, 1)$ then the equality of these probabilities 
is equivalent to $f_1(\vec{s})=f_2(\vec{1}).$ Similarly, replacing
$n$ with $-n$ gives $f_1(-\vec{s})=f_2(-\vec{1})$ and
$$
f_1(\vec{s})\bigr/f_1(-\vec{s})
=
f_2(\vec{1})\bigr/ f_2(-\vec{1}).
$$
This implies that $h\sum \mbox{sgn}(a_l) = h n.$ If $h\not = 0$
then all $a_l=1$ that contradicts that $u<0$ 
for which $a_l= \xi'(s^2)\xi'(u)<0.$
\qed

We are ready to finish the proof of Theorem \ref{Thmain}.
Under the conditions of Lemma \ref{L4}, the inequality in (\ref{Psipsi})
will be strict for all $\vec{\lambda}.$ However, in the computation leading
from (\ref{Psipsi}) to (\ref{big}) we only really need to use this 
(strict) inequality for $\vec{\lambda}_0$ such that
\begin{equation}
\psi(Q,\vec{\lambda}_0)= \inf_{\vec{\lambda}}\psi(Q,\vec{\lambda})
\label{inf}
\end{equation}
By convexity, such $\vec{\lambda}_0$ exists and is unique.
Then the same computation gives that
\begin{equation}
\inf_{\vec{\gamma}}\Psi(P,\vec{\gamma}) 
< n\, \RS(|\vec{u}|_p)\leq \P(a)
\label{big2}
\end{equation}
and Lemma \ref{L2} finishes the proof of Theorem.
One small technical issue that needs to be mentioned is that
if some $w_{l,l'}=\pm 1$ then corresponding $\lambda_{l,l'}\to\pm \infty$
in the infimum (\ref{inf}).
However, this does not cause a problem, it simply forces the corresponding
coordinates in $\Phi_n$ to 'glue' together by forcing $\eps_l = \eps_{l'},$
but the rest of the argument remains the same.

\qed

\end{document}

%% file: invlat.tex
\font\tencmmib=cmmib10 \skewchar\tencmmib '60
\newfam\cmmibfam
\textfont\cmmibfam=\tencmmib

\def\bbox{\quad\hbox{\vrule \vbox{\hrule \vskip2pt \hbox{\hskip2pt
\vbox{\hsize=1pt}\hskip2pt} \vskip2pt\hrule}\vrule}}
\def\lessim{\ \lower4pt\hbox{$
\buildrel{\displaystyle <}\over\sim$}\ }
\def\gessim{\ \lower4pt\hbox{$\buildrel{\displaystyle >}
\over\sim$}\ }

\def\eps{\varepsilon}

\def\go0{\to 0}

\def\la{\langle}

\def\leftitem#1{\item{\hbox to\parindent{\enspace#1\hfill}}}

\def\qed{\hfill\break\rightline{$\bbox$}}

\def\ra{\rangle}

\def\sg{\sigma}

\def\sg2{\sigma^2}

\def\__{_{\infty}}